\documentclass[twocolumn,preprintnumbers,amsmath,amssymb,showpacs,prb,aps]{revtex4-1}
\usepackage{graphicx}
\usepackage{dcolumn}
\usepackage{bm}
\usepackage{SIunits}
\usepackage{verbatim}
\usepackage{placeins}

\begin{document}
\title{Interlayer magnetic frustration driven quantum spin disorder
      in honeycomb compound In$_{3}$Cu$_{2}$VO$_{9}$}
\author{Da-Yong Liu$^{1}$, Ying Guo$^{1}$, Xiao-Li Zhang$^{1}$, Jiang-Long Wang$^{2}$,
        Zhi Zeng$^{1}$, H. Q. Lin$^{3}$ and Liang-Jian Zou$^{1}$ }
\altaffiliation{Corresponding author} \email{zou@theory.issp.ac.cn}
\affiliation{
      $^1$ Key Laboratory of Materials Physics, Institute of Solid State
      Physics, Chinese Academy of Sciences, P. O. Box 1129, Hefei, Anhui
      230031, China \\
      $^2$ College of Physics Science and Technology, Hebei University,
      Baoding 071002, China \\
      $^3$ Department of Physics, Chinese University of Hong Kong,
      Shatin, New Territory, Hong Kong, China}

\date{\today}

\begin{abstract}
We present electronic and magnetic properties of a
honeycomb compound In$_{3}$Cu$_{2}$VO$_{9}$ in this paper. We find
that the parent phase is a charge transfer insulator with an energy
gap of about 1.5 eV. Singly occupied d$_{3z^{2}-r^{2}}$ electrons of
copper ions contribute an $S$ = 1/2 spin, while vanadium ions show nonmagnetism.
Oxygen 2$p$ orbitals hybridizing with a small fraction of Cu 3$d$ orbitals
dominate the density of states near $E_{F}$. The planar nearest-neighbor,
next-nearest-neighbor and interplane superexchange couplings
of Cu spins are $J_{1}$ $\approx$ 16.2 meV, $J_{2}$ $\approx$ 0.3 meV and
$J_{z}$ $\approx$ 1.2 meV, suggesting a low-dimensional
antiferromagnet \cite{Sondhi10}. We propose that the magnetic
frustration along the $c$-axis leads to a quantum spin disorder in
In$_{3}$Cu$_{2}$VO$_{9}$, in accordance with the recent experiments.
\end{abstract}


\vskip 300 pt

\maketitle

\section{Introduction}

   Atoms in reduced dimensional lattice, for example, in two-dimensional (2D)
honeycomb lattice, has low coordinate number and small spin number. Together
with magnetic frustration, these electrons
may experience strong spin fluctuations and probably form extremely
low-dimensional (low-D) antiferromagnet (AFM), or even exotic spin liquid,
quantum disorder, or quantum spin Hall state, so it has
attracted great interest in recent theories
\cite{Meng10, White10, PALee07, WenXG10, Sondhi10, Wangfa10, Liebsch10, Kou11,
Ran10, XiangT11, WangQH11, LiuWM11, LiTao12}.
A recent synthesized honeycomb compound In$_{3}$Cu$_{2}$VO$_{9}$
\cite{Moller08,Yehia10,Yan11} is such a system with reduced dimensionality
and frustration.
According to the chemical valence analysis, the copper ions with 3$d^9$
configuration contribute a spin-1/2 magnetic moment, while the 3$d$ orbitals
in vanadium ions are empty and contribute no magnetic moment. Due to the
large separation between Cu/V-O layers, the copper spins form a
quasi-2D honeycomb lattice \cite{Yehia10}. This compound exhibits
unusual magnetic behavior: it shows neither an AFM long-range order in magnetic
susceptibility measurement, nor any magnetic phase transition peak in specific
heat data \cite{Moller08,Yan11}.
However, Zn or Co doping can induce a weak or strong three-dimensional (3D)
AFM long-range order observed experimentally \cite{Yan11}.
It inspires great interest in such a topic: whether the ground state of this
compound is low-dimensional AFM, or spin liquid predicted by Meng {\it et al.} \cite{Meng10}.

Additionally, it is well known that noninteracting or weakly interacting 
electrons in hexagonal graphene is a typical example of {\it Dirac} fermions.
It is not very clear that how the properties of Dirac fermions evolve
with the increase of Coulomb interactions.
Therefore, In$_{3}$Cu$_{2}$VO$_{9}$ with hexagonal Cu ions is probably a
realistic material to testify how the electronic correlations
affect the {\it Dirac} fermions.
In the meantime, through the magnetic susceptibility and specific
heat measurements, it's hard to distinguish possible ground state between
a low-dimensional AFM and a spin liquid \cite{Yan11}. All of these drive us to explore
the electronic properties and magnetism in In$_{3}$Cu$_{2}$VO$_{9}$ by
combining the first-principles electronic structure calculations
and analytic methods, since it could provide us first
insight to this compound.

Utilizing the local density functional (LDA) and
the correlation correction approaches, together with the analytical
perturbation method, we show that In$_{3}$Cu$_{2}$VO$_{9}$ is an
AFM charge transfer insulator with copper d$_{3z^{2}-r^{2}}$ orbitals half
filled and contributing $S$ = 1/2 spins.
However, vanadium ions are nonmagnetic (NM). The oxygen $2p$ orbitals
dominate the density of states (DOS) near $E_{F}$. We extract not only
tight-binding parameters between copper ions, but also various spin
coupling strengths between Cu spins. Furthermore, we show that the absence
of 3D AFM long-range order observed experimentally in In$_{3}$Cu$_{2}$VO$_{9}$
arises from the magnetic frustration of Cu spins
along the $c$-axis, leading to a quantum disorder phase.
Meanwhile, in terms of this new scenario, we address the evolution of the
magnetic properties on the Co and Zn dopings in In$_{3}$Cu$_{2}$VO$_{9}$,
giving a plausible explanation of the experimental behavior.
We also suggest the possibility that doping in oxygen
sites may drive the system to a spin liquid or superconducting phase.
The rest of this paper is organized as follows: in Sec. II, we briefly
describe the unit cell of In$_{3}$Cu$_{2}$VO$_{9}$ and numerical
calculation methods; in Sec. III, we present
the major numerical results and our analysis; Sec. IV is devoted to
the remarks and conclusion.

\section{Numerical Methods}

We first study the electronic state properties of the honeycomb compound
In$_{3}$Cu$_{2}$VO$_{9}$ by employing the first-principles
electronic structure calculation approach so as to elucidate its
groundstate properties.
The electronic structure calculations were performed using the
self-consistent full potential linearized augmented-plane-wave (FP-LAPW)
scheme in the WIEN2K programme package.
\cite{WIEN2K}. We used 36 $k$ points in the irreducible part of the
first Brillouin zone. The muffin-tin sphere radii were selected to
be 2.36, 1.93, and 1.71 a.u. for In, Cu (V), and O in
In$_{3}$Cu$_{2}$VO$_{9}$. The plane-wave cutoff parameter $R_{MT}K_{max}$
was 7.0, and the cutoff between the core and valence states was $-$7.0
Ry in all calculations. Exchange and correlation effects were taken
into account in the generalized gradient approximation (GGA) by
Perdew, Burk, and Ernzerhof (PBE) \cite{PBE}. In order to explicitly
take into account the correlated effect of the 3$d$ electrons of Cu (V)
ions, we also performed GGA+$U$ calculations for
In$_{3}$Cu$_{2}$VO$_{9}$. The effective Coulomb interaction
$U_{eff}$ = $U-J_{H}$ ($U$ and $J_{H}$ are the on-site Coulomb
interaction and the Hund's rule coupling, respectively) was used
instead of $U$. We take $U_{eff}$ = 8.0 eV for copper and 4.0 eV for
vanadium according to the similar compounds
\cite{PRB79-125201,PRL104-047401,JPCM22-416002}.

\vspace{1mm}
\begin{figure}[htbp]
\centering
\includegraphics[angle=0, width=0.8 \columnwidth]{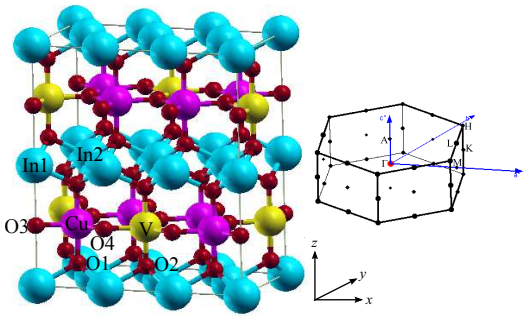}
\caption{(Color online) (a) Unit cell of In$_{3}$Cu$_{2}$VO$_{9}$
($Cmcm$ space group), and corresponding high symmetry points
in the first Brillioune zone.
} \label{fig1}
\end{figure}
To model the low-energy process in In$_{3}$Cu$_{2}$VO$_{9}$, the tight-binding
parameters in the NM (or paramagnetic (PM)) phase are necessary. We also perform
the local density functional (LDA) calculations for NM
In$_{3}$Cu$_{2}$VO$_{9}$. The NM band structures of
In$_{3}$Cu$_{2}$VO$_{9}$ are obtained by using the full-potential linearized
augmented plane-wave plus local orbitals (FP-LAPW+lo) scheme
implemented in the WIEN2K package \cite{WIEN2K}.
In order to compare our numerical results with the experimental data,
we adopt the experimental structural data of In$_{3}$Cu$_{2}$VO$_{9}$
measured by neutron powder diffraction \cite{Moller08}. This compound has an
orthorhombic structure with space group $Cmcm$ and lattice constants
$a$ = 10.0491 ${\AA}$, $b$ = 5.8019 ${\AA}$, and $c$ = 11.8972 ${\AA}$, as shown in
Fig. 1.
It was found that In$_{3}$Cu$_{2}$VO$_{9}$ has a
honeycomb layered structure with Cu atoms forming hexagonal net and
V atoms in the center of an hexagon. A Cu atom is surrounded by five O atoms
in a trigonal bipyramidal environment. Detailed results of band structure are
addressed in what follows.

\subsection{Nonmagnetic State}

We first present the NM electronic
structures of In$_{3}$Cu$_{2}$VO$_{9}$ in Fig. 2. It shows that the major
orbital character of the band structures near $E_F$ is a
${3z^2-r^2}$ symmetry of copper 3$d$ orbitals. This Cu 3$d_{3z^2-r^2}$ orbital
in the honeycomb lattice, similar to the $p_{z}$ orbital of carbon in
graphene, contributes an outstanding property of the
band structures, {\it i.e.}, a {\it Dirac} cone with the approximately
linear spectrum is observed around the $H$ point, as seen in Fig. 2.
The linear energy spectrum ranges from $-$0.2 to 0.2 eV. We
notice that the original {\it Dirac} point around the $K$ points opens a
small energy gap, about 0.2 eV, also as seen
in Fig. 2, which is attributed to the weak interlayer coupling
between Cu/V-O hexagonal planes.
%
\begin{figure}[htbp]
\centering
\includegraphics[angle=0, width=0.8 \columnwidth]{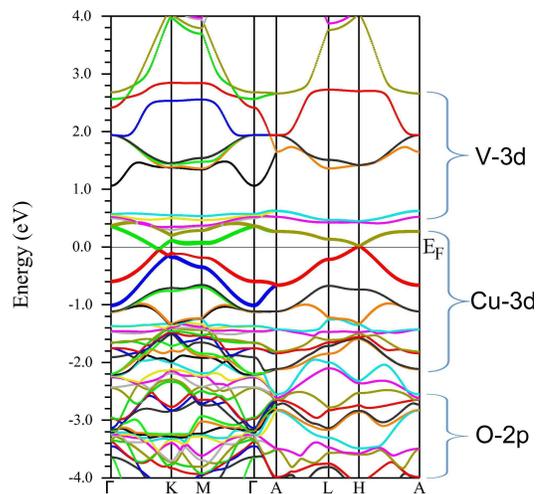}
\caption{(Color online) Band structures with orbital
character of In$_{3}$Cu$_{2}$VO$_{9}$ in the NM situation
by the LDA method. We indicate the major orbital component of each band. 
The bands marked with
heavy lines are Cu 3$d_{3z^2-r^2}$ orbitals.} \label{fig2}
\end{figure}
Though parts of empty V 3$d$ orbital are near the top of Cu 3$d_{3z^2-r^2}$
orbitals, they do not directly hybridize with each other considerably.
A part of filled oxygen 2$p$ orbitals are close to the bottom of Cu
3$d_{3z^2-r^2}$ orbitals, and these 2$p$ orbitals also do not considerably mix
with 3$d_{3z^2-r^2}$ orbitals, as we see the partial density of states (PDOS)
of In$_{3}$Cu$_{2}$VO$_{9}$ in Fig. 3. Thus the
low-energy physics in In$_{3}$Cu$_{2}$VO$_{9}$ could be
approximately described by the Cu 3$d_{3z^2-r^2}$ orbitals across$E_{F}$.
\begin{figure}[htbp]
\centering
\includegraphics[angle=0, width=0.95 \columnwidth]{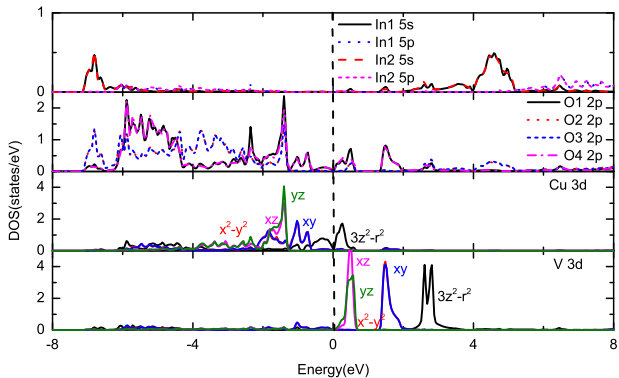}
\caption{(Color online) Partial density of states (PDOS) of
In$_{3}$Cu$_{2}$VO$_{9}$ in the NM situation by the LDA method.
} \label{fig3}
\end{figure}

Based on the fact that Cu 3$d_{3z^2-r^2}$ orbitals mainly distribute
from $-$1.0 eV to 0.4 eV, we extract these four Cu 3$d_{3z^2-r^2}$
bands (notice that we have four Cu atoms in primitive cell) from the full band
structures and fit these bands with a
single-orbital tight-binding model with the nearest-neighbor (NN),
next NN (NNN) and $3rd$ NN hopping integrals in the Cu/V-O plane and
between the planes. These tight-binding parameters are listed in
Table I. Notice that the LDA methods deal with the NM state
rather than the PM one. But as we know, the overlap and hopping
integrals of electrons depend on the spatial wavefunctions, and
are independent of the spin degree of freedom. Thus the hopping
parameters obtained in the NM state are the same to those
in the PM state.
\begin{table}[b]
\caption{Tight-binding parameters $t_{\alpha}$ and superexchange spin
coupling strengths $J_{\alpha}$ in In$_{3}$Cu$_{2}$VO$_{9}$. Subscripts
${1,2,3}$ and ${c}$ represent the NN, NNN, $3rd$ NN hopping
integrals or spin coupling strengths in the planes, and
the NN ones between the planes, respectively. All the energies are measured
in units of meV.}
\label{momreal}
\begin{center}
\begin{tabular}{lrccccccccccc}
\hline\hline
$t_{1}$ & $t_{2}$ & $t_{3}$ & $t_{c}$ &  $J_{1}$ & $J_{2}$ & $J_{c}$  \\
\hline -180.2 & -24.1 & 5.8 & -48.9 &  16.2 & 0.3 & 1.2\\
\hline\hline
\end{tabular}
\end{center}
\end{table}
One can easily find that the intraplane NN hopping integral $t_{1}$ is
about 7.5, 31 and 3.7 times of the intraplane NNN hopping $t_{2}$, 3$rd$
hopping $t_{3}$ and the interplane NN hopping $t_{c}$, respectively,
suggesting the NN hopping and NN spin coupling dominate the transfer process
and magnetic correlations in the ground state of In$_{3}$Cu$_{2}$VO$_{9}$.

\subsection{Degenerate Antiferromagnetic States}

To understand the magnetism and insulating nature of undoped compound,
we further adopt the spin-polarized GGA and correlation-corrected GGA+$U$
schemes to uncover the groundstate properties of
strongly correlated In$_{3}$Cu$_{2}$VO$_{9}$. We consider five kinds of
different magnetic configurations, NM, ferromagnet (FM), A-type AFM
({\it i.e.}, interlayer AFM and intralayer FM), and
G1- and G2-type AFM as the possible candidate ground state. 
Here G1- and G2-type AFM configurations shown in Fig. 6 are two different 
magnetic structures, both of which are derived from a conventional 
G-type AFM. In comparison with the G1-type AFM structure, the lower Cu/V-O 
hexagon (or Cu/V-O layer) in the G2-type structure rotates $\pi$/3 degree. 
Thus the G1-type spin configuration differs from the G2-type one.  
Our GGA+$U$ numerical results show that
the total energy of the NM state is the highest, and that
of the G1- and G2-type AFM states are the lowest. Interestingly, we find that
the G1- and G2-type AFM states are degenerate. This arises from the interlayer
magnetic frustration of Cu spins, as will be discussed in detail later. 
The total energy difference between
the FM and A-type AFM configurations is rather small.
The energy difference between the lowest G1(G2)-type AFM and the second
lowest A-type AFM or FM states is about 58.5 meV, suggesting that the AFM
magnetic couplings between Cu spins in In$_{3}$Cu$_{2}$VO$_{9}$ are dominant.

To examine the effect of the on-site Coulomb interaction parameter $U$ on the
calculation results, we also perform the spin-polarized GGA calculations for
the purpose of comparison. The PDOS of In$_{3}$Cu$_{2}$VO$_{9}$ for G1(G2)-type
AFM state is shown in Fig. 4.
\begin{figure}[htbp]
\centering
\includegraphics[angle=0, width=0.95 \columnwidth]{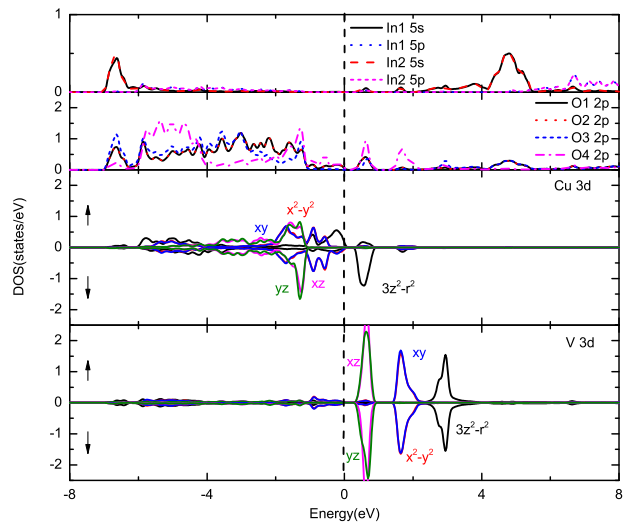}
\caption{(Color online) PDOS of In$_{3}$Cu$_{2}$VO$_{9}$ for G1(G2)-type AFM state
by the spin-polarized GGA method.
} \label{fig4}
\end{figure}
It is found that the system is an insulator with an energy gap of 0.4 eV. However the
calculated magnetic moment is only 0.4 $\mu_{B}$, smaller than
that observed by the experiment, about 0.5 $\mu_{B}$ \cite{Moller08}.
Considering the fact that the LSDA results usually overestimate the magnitude of
a magnetic moment, while the experimental results underestimate it due to the
significant disorder and fluctuations in realistic compounds, one can expect
that there exists a larger magnetic moment than 0.5 $\mu_{B}$.
In a similar system Cu$_{2}$V$_{2}$O$_{7}$ with
0.73 $\mu_{B}$ for $U_{eff}$= 6.52 eV \cite{PRB79-125201}, the GGA+$U$ calculations
are also needed for In$_{3}$Cu$_{2}$VO$_{9}$. We choose the Coulomb parameters
lying in a physical parameter range for In$_{3}$Cu$_{2}$VO$_{9}$, {\it i.e.}
$U_{eff}$= 8 eV for Cu ions, and 4 eV for V ions according to the similar compounds,
such as Cu$_{2}$V$_{2}$O$_{7}$ \cite{PRB79-125201} and
In$_{2}$VO$_{5}$ \cite{JPCM22-416002}, as well as the general compounds,
such as cuprates and vanadium oxides (V$_{2}$O$_{3}$ \cite{PRL104-047401}), {\it etc.}.
Notice that we perform the numerical calculations for several different
Coulomb parameters for Cu and V atoms to systematically investigate their
influence on our results, {\it i.e.} $U_{eff}$= 6 $\sim$ 8 eV for Cu \cite{PRB79-125201}
and $U_{eff}$= 3 $\sim$ 5 eV for V \cite{JPCM22-416002,PRL104-047401}.
From our numerical calculations, we find that the band gap and the magnetic 
moment vary from 1.37 to 1.5 eV and from 0.72 to 0.79 $\mu_{B}$, respectively, 
when U$_{eff}$ changes from 6 to 8 eV for Cu and from 3 to 5 eV for V. This 
clearly shows that the electronic and magnetic properties do not crucially 
depend on the values of the parameter U in the physical range.

To further uncover the insulating character of In$_{3}$Cu$_{2}$VO$_{9}$,
the PDOS for G1(or G2)-type AFM state by the GGA+$U$ approach is
plotted, as seen in Fig. 5. The system is obviously an insulator with an
energy gap of 1.5 eV.
We found that oxygen 2$p$ orbitals consist of the major bands near $E_{F}$,
ranging from 0 to $-$4.0 eV. While the gravity center of Cu 3$d$ orbitals lie far from
$E_{F}$. The center of the spin-up 3$d$ orbital lies
at about $-$5.9 eV, and that of the spin-down 3$d$ orbitals lies at about
$-$4.8 eV. This gives rise to the exchange splitting of Cu 3$d$ orbitals
about $\Delta_{ex}$ $\approx$ 1.1 eV, implying that the Hund's rule coupling
$J_{H}$ = $\Delta_{ex}/2S$ $\approx$ 1.1 eV. Such a large exchange splitting
also shows a strong Coulomb correlation on Cu 3$d$ orbitals.
Note that the charge transfer energy between O 2$p$ orbital and Cu 3$d$
orbital in In$_{3}$Cu$_{2}$VO$_{9}$, $\Delta$ = $\epsilon_{2p}-\epsilon_{3d}$,
which has the same order as the insulating energy gap, is considerably smaller
than the on-site Coulomb interaction of Cu 3$d$ electrons, demonstrating that
In$_{3}$Cu$_{2}$VO$_{9}$ is a charge transfer insulator.
\begin{figure}[htbp]
\centering
\includegraphics[angle=0, width=0.95 \columnwidth]{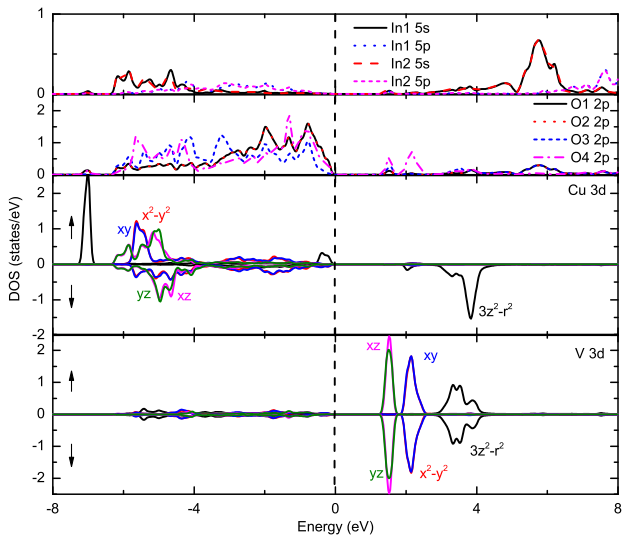}
\caption{(Color online) PDOS of In$_{3}$Cu$_{2}$VO$_{9}$ for G1(G2)-type AFM state
by the spin-polarized GGA+$U$ method.
} \label{fig5}
\end{figure}

The charge distribution and occupation numbers in the 3$d$ orbitals of
Cu and V are essential to understand the electronic properties of
In$_{3}$Cu$_{2}$VO$_{9}$. Our results show that in the Cu 3$d$
orbitals, the hole mainly occupies the active $d_{3z^{2}-r^{2}}$
orbital. It has a total number of 1.17 electrons, or 0.83 holes, in
the $d_{3z^{2}-r^{2}}$ orbital. In contrary to the chemical valence analysis,
V has totally 1.87 electrons in its 3$d$ orbitals, though it shows
NM or no spin polarization in the present GGA+$U$ ansatz.
This arises from a strong $p$-$d$ hybridization between V and O atoms, and a
fraction of electrons transfer from oxygen 2$p$ orbitals to vanadium
3$d$ orbitals.

Fig. 5 displays that the main DOS of copper lies from $-$7 eV to $-$4 eV,
though a tiny portion of 3$d$ electrons lie between $-$4 eV to $E_{F}$
which arises mainly from the hybridization of Cu 3$d$ orbital with the 2$p$
orbitals near $E_{F}$ of surrounding O anions. It's observed from Fig. 5 that the
3$d_{3z^{2}-r^{2}}$ orbital in Cu is almost fully spin polarized.
The magnetic moment of Cu spin is almost contributed from the well
localized spin-up 3$d_{3z^{2}-r^{2}}$ orbital at the position of
about 7 eV below $E_{F}$. We find that the orbital ground state is the
Cu 3$d_{3z^{2}-r^{2}}$ character, in agreement with the electron spin resonance
experiment by Kataev {\it et al.} \cite{Kataev05}.

Meanwhile, a small but finite DOS distributed from $-$6 eV to $E_{F}$ is
seen in V 3$d$ orbitals, showing that the two 3$d$ electrons form a wide
energy band below the Fermi energy. Obviously, other V 3$d$ electrons
contribute the highest empty conduction bands, as we see in Fig. 5. The
PDOS of V shows that two 3$d$ electrons in V are not spin polarized and
allocate in five 3$d$ orbitals with almost equal weight, in sharp contrast
to those of Cu atoms. Fig. 5 shows that In 5$p$ orbitals give rise to a small
DOS in a wide energy range. In contrast, the DOS of oxygen mainly distributes
near $E_{F}$ and is greatly larger than that of In.
Such a strongly correlated character in In$_{3}$Cu$_{2}$VO$_{9}$ is
similar to that in the parent phases of high-T$_{c}$ cuprates. This
implies that the doping in O sites can readily affect the transport
properties in In$_{3}$Cu$_{2}$VO$_{9}$, and easily drive the
system transit to metallic or superconducting phase once hole carriers are doped
in oxygen 2$p$ bands.

\subsection{Magnetic Coupling Strengths}

The groundstate magnetic properties of In$_{3}$Cu$_{2}$VO$_{9}$ is
the central issue of this paper. As described above, our
numerical results demonstrate that the stablest phase is the degenerate
G1- and G2-type AFM states. The magnetic moment mainly contributes from
copper spins with a total net magnetic moment of 0.79 $\mu_{B}$ per Cu
atom in which the average electron number is 0.98 in the spin-up band, and 0.19
in the spin-down one. The magnetic moment mainly arises from the
$d_{3z^{2}-r^{2}}$ orbital of copper. Apparently, a strong
electron correlation and a large Hund's rule coupling result in such a
large spin polarization, leading to a well defined local moment in
coppers. In contrast, vanadium is NM, though it is 3$d^{2}$
configuration with two electrons distributing from $E_{F}$ to $-$7.0 eV.
Its NM character may arise from the strong {\it p-d} hybridization
between V and O, resulting in a weak correlation between V 3$d$ electrons.
We do not rule out the possibility of spin frustrations in the present Cu-V
geometry, which may deserve further discussion in the future. Due to the full
fillings of the oxygen 2$p$ orbitals and In 5$p$ orbitals, both O and In are
not spin polarized.

The insulating nature in In$_{3}$Cu$_{2}$VO$_{9}$
suggests that the local spins of coppers interact through a
superexchange couplings mediated via oxygens and vanadium. One
expects that the contributions to the NN and NNN Cu-Cu
superexchange couplings arise from the direct hopping between
$d_{3z^2-r^2}$ electrons, the indirect hopping between Cu spins
through the $p$-$d$ hybridization of intermediate oxygen anions and
through the O-V-O bridge. To quantitatively
obtain the superxchange coupling strengths, we utilize the
tight-binding parameters obtained in Table I to estimate the NN and
NNN superexchange couplings between Cu spins in the Cu/V-O plane and
between the planes, taking the effective Coulomb interaction $U$ as 8
eV. A $t$/$U$ expansion of the Hubbard model is used to
obtain an effective $J_{1}$-$J_{2}$-$J_{c}$ spin $S$=1/2 model on a honeycomb
lattice:
\begin{eqnarray}
   H &=&
J_{1}\sum_{\substack{<ij>_{ab}}}\overrightarrow{S}_{i}\cdot\overrightarrow{S}_{j}
+J_{2}\sum_{\substack{<<ij>>_{ab}}}\overrightarrow{S}_{i}\cdot\overrightarrow{S}_{j}
\nonumber \\
&&+J_{c}\sum_{\substack{<ij>_{c}}}\overrightarrow{S}_{i}\cdot\overrightarrow{S}_{j},
\end{eqnarray}
where $J_{1} = 4t_{1}^{2}/U-16t_{1}^{4}/U^{3}$, $J_{2} = 4t_{2}^{2}/U+4t_{1}^{4}/U^{3}$ and $J_{c} = 4t_{c}^{2}/U$ are the NN, NNN
intralayer and the NN interlayer magnetic couplings, respectively.
We find that the intraplane and interplane NN superexchange
couplings are about $J_{1} \approx$ 16.2 meV and $J_{c} \approx$ 1.2
meV, respectively; and the intraplane NNN superexchange couplings
are about $J_{2} \approx$ 0.3 meV. More long-range spin couplings
are so small to be negligible. This gives rise to $J_{2}$/$J_{1}$
$\approx$ 0.02, in addition to a rather weak interlayer coupling of
$J_{c}$/$J_{1}$ $\approx$ 0.07. All of these spin coupling strengths have
been listed in Table I. 
From the literature available, we find that the spin coupling strengths of 
In$_{3}$Cu$_{2}$VO$_{9}$ fall in the parameter range of a 2D Heisenberg 
AFM \cite{Sondhi10,Mezzacapo12}, showing that In$_{3}$Cu$_{2}$VO$_{9}$
is an insulator with 2D AFM. As we will illustrate later, such a weak 
interlayer magnetic coupling and the spin frustration prevent from the 
long-range AFM order in In$_{3}$Cu$_{2}$VO$_{9}$.

\section{Remarks and Conclusions}

It is observed experimentally that neither the specific heat nor magnetic
susceptibility exhibits AFM-PM transition
\cite{Yan11}: the former is almost T$^2$ dependent, and the
latter linearly increases with the lift of
temperature. Consequently, one naturally suspects that the ground state of
In$_{3}$Cu$_{2}$VO$_{9}$ might be a gapless spin liquid phase, a
long-time searched exotic quantum phase both theoretically and
experimentally, since such a phase in {\it Kagome} lattice
also demonstrates similar unusual temperature-dependent behavior
in the low T regime \cite{Ran07,PALee07}. Meanwhile, a
strongly low-D AFM may exhibit similar behaviors \cite{low-D-AFM}.
Though both a low-dimensional AFM and a spin liquid have local
permanent magnetic moments, and do not display any sign of ordering
in specific heat and magnetic susceptibility down to the lowest
temperatures despite of comparable strong AFM interactions. A low-dimensional
AFM could be different from a spin liquid in many aspects. For example, a 2D AFM
has a long-range correlation and long-range order at T= 0 K as well as
short-range one at finite T, while a spin liquid only has short-range
correlation, even at T= 0 K; this leads to distinct low-temperature behavior 
in magnetic susceptibility for a 2D AFM and a spin liquid, {\it etc.}.
Therefore our theoretical analysis above suggests
2D AFM ground state rather than a spin liquid, though it is still
mysterious why In$_{3}$Cu$_{2}$VO$_{9}$ exhibits neither an
AFM long-range order in the low temperature regime nor a
Curie-Weiss law in the high temperature regime \cite{Yan11,Moller08}, which is
exotic and distinctly different from undoped cuprates.

An earlier experimental study by M$\ddot{o}$ller {\it et al.}
\cite{Moller08} attributed the absence of a 3D AFM long-range order
to the disorder stacking of 2D Cu$/$V order domains along the $c$-axis.
However, from our preceding data and the particular geometry of In$_{3}$Cu$_{2}$VO$_{9}$,
as an alternative explanation, we propose that it is the strong AFM frustration
along the $c$-axis leads to the absence of the 3D AFM long-range order
in In$_{3}$Cu$_{2}$VO$_{9}$. As seen the Cu-V geometric configuration among
two Cu-V layers in Fig. 6,
\begin{figure}[htbp]
\centering
\includegraphics[angle=0, width=0.8 \columnwidth]{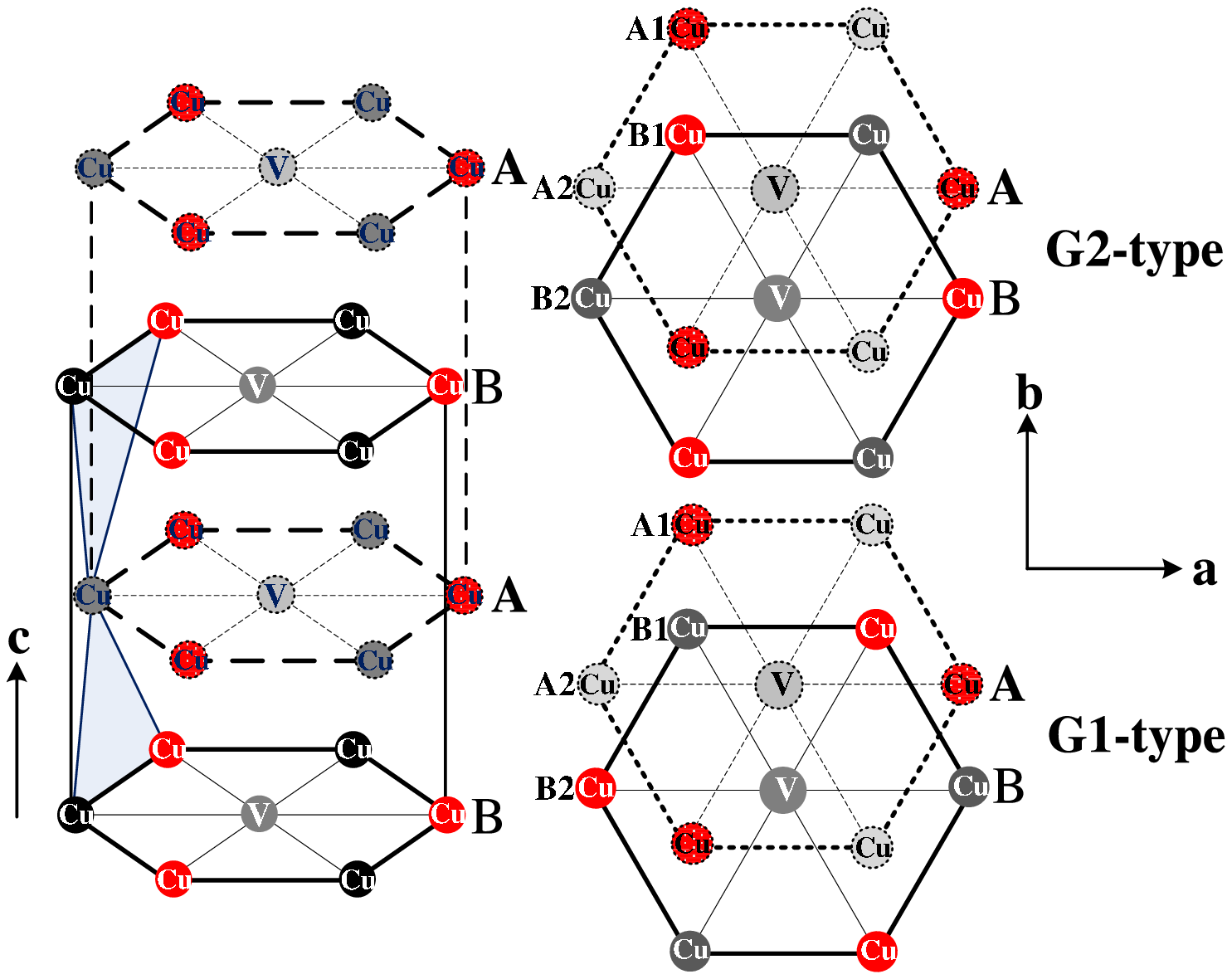}
\caption{(Color online) Schematics of frustrated magnetic configurations of
Cu spins between A and B layers (left panel), and a top view of the G1(G2)-type AFM
(right panel). A Cu spin in the upper layer (dash line) has the same
distance to two Cu spins in the lower layer (solid line), consisting of an isosceles
triangle. Filled gray and red circles represent spin down and spin up, respectively.} \label{fig6}
\end{figure}
a Cu spin has two NN layers along the $c$-axis, it has the same
distance to other two Cu spins in the upper or lower layer, forming an isosceles
triangle which is a notorious spin configuration with magnetic frustration.
Our analysis above has shown that both the spin couplings of the interlayer and
intralayer spins are AFM, indicating the spin fluctuations along the
$c$-axis will be very large. Thus in In$_{3}$Cu$_{2}$VO$_{9}$ the intralayer
Cu spins are AFM correlated, while the interlayer Cu spins are quantum disordered.
Such a frustration scenario could unifiedly account for both the presence of
low-D AFM correlation \cite{Yehia10} and the absence of
3D AFM order \cite{Kataev05,Moller08,Yan11} in recent experiments.
\begin{figure}[htbp]
\centering
\includegraphics[angle=0, width=0.8 \columnwidth]{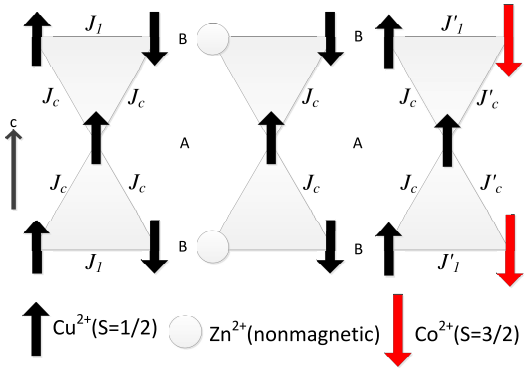}
\caption{(Color online) Schematics of the frustration scenario to explain
the evolution of the magnetic behavior in Zn- and Co- doped In$_{3}$Cu$_{2}$VO$_{9}$.
} \label{fig7}
\end{figure}

Furthermore, we apply our frustration scenario to uncover the
evolution of the magnetic properties on the Co and Zn dopings
in In$_{3}$Cu$_{2}$VO$_{9}$, giving a possible explanation of the experimental
behavior, as shown in Fig. 7. The Zn and Co dopings break the
inter-layer frustration, while the NM Zn$^{2+}$ ions destroy the
intra-layer AFM. On the contrary, high spin-3/2 Co$^{2+}$ ions enhance it.
Thus the Zn doping induces a weak AFM at low doping, but suppresses the AFM
state completely at high doping. In comparison, the Co doping results in a
strong AFM with large effective magnetic moment. All the results are in
accordance with the experimental observations.

Moreover, as we point out above, the hole doping in O sites of
In$_{3}$Cu$_{2}$VO$_{9}$ may introduce carriers near Fermi energy.
With the increase of carriers, one expects that the screening effect can
greatly reduce the Coulomb interaction, and weaken the electronic
correlation. This may drive the In$_{3}$Cu$_{2}$VO$_{9}$ system
from an AFM ground state to a spin liquid phase \cite{Meng10},
since in the intermediate correlation regime, a half-filled Hubbard system
with the honeycomb structure exists a spin-liquid ground state
\cite{Meng10,WenXG10}.
Nevertheless, we don't expect that doped In$_{3}$Cu$_{2}$VO$_{9}$ will
form the well-known Zhang-Rice singlet in the Cu-O plane of
cuprates, since the latter arises from the hybridization between Cu
3$d_{x^{2}-y^{2}}$ and O 2$p$ orbitals, forming delocalized states in the
xy plane; while in the former, the 3$d_{3z^{2}-r^{2}}$ orbital mixes with
O 2$p$ orbital, forming a delocalized state in the $z$-direction. This results
in completely different transport properties in doped compounds, as Yan
{\it et al.} observed in In$_{3}$Cu$_{2-x}$Co$_{x}$VO$_{9}$ and
In$_{3}$Cu$_{2-x}$Zn$_{x}$VO$_{9}$ \cite{Yan11}.

In summary, the investigation of the electronic structure and magnetic
properties in quasi-2D honeycomb compound
In$_{3}$Cu$_{2}$VO$_{9}$ demonstrates that the undoped
phase is a charge transfer insulator with a gap
of 1.5 eV , and oxygen 2$p$ orbitals dominate the bands near $E_{F}$.
Spin-1/2 local moment antiferromagnetically interacts and occupies
the 3$z^{2}$-$r^{2}$ orbital in copper. In contrast, the NM V ions
have two electrons in the 3$d$ orbitals. The
tight-binding parameters and estimated spin coupling strengths
suggest that In$_{3}$Cu$_{2}$VO$_{9}$ is a strongly
2D {\it N\'{e}el} AFM. We propose that the spin fluctuations
arising from the magnetic frustration along the $c$-axis destroy the 3D
AFM long-range order, leading to random stacking of
Cu-V layers.
\\

\acknowledgements

We acknowledge X. H. Chen provides us with original experimental data.
This work was supported by the National Sciences Foundation of China
under Grant No. 11074257 and 11104274, and Knowledge Innovation Program
of the Chinese Academy of Sciences. Numerical calculations were
performed at the Center for Computational Science of CASHIPS.


\begin{references}

\bibitem{Meng10} Z. Y. Meng, T. C. Lang, S. Wessel, F. F. Assaad, and A.
Muramatsu, Nature {\bf 464}, 847 (2010).

\bibitem{White10} S. M. Yan, D. A. Huse, and S. W. White, Science {\bf 332},
1173 (2011).

\bibitem{PALee07} S. Saremi and P. A. Lee, Phys. Rev. B {\bf 75}, 165110 (2007).

\bibitem{WenXG10} A. Vaezi and X.-G. Wen, arXiv:1010.5744; arXiv:1101.1662.

\bibitem{Sondhi10} B. K. Clark, D.A.Abanin, and S. L. S
ondhi, Phys. Rev. Lett. {\bf 107}, 087204 (2011).

\bibitem{Wangfa10} F. Wang, Phys. Rev. {\bf B 82}, 024419 (2010).

\bibitem{Liebsch10} A. Liebsh, Phy. Rev. B {\bf 83}, 035113 (2011).

\bibitem{Kou11} J. He, S.-P. Kou, Y. Liang and S.-P. Feng, Phys. Rev. {\bf B 83}
205116 (2011); {\it ibid}, {\bf 84}, 035127 (2011).

\bibitem{Ran10} Y.-M. Lu and Y. Ran, Phys. Rev. B {\bf 84}, 024420 (2011).

\bibitem{XiangT11} H. H. Zhao, Q. N. Chen, Z. C. Cai, M. P. Qin, G. M. Zhang,
and T. Xiang, arXiv:1105.2716.

\bibitem{WangQH11} W.-S. Wang, Y.-Y. Xiang, Q.-H. Wang, F. Wang, F. Yang, and
D.-H. Lee, Phys. Rev. B {\bf 85}, 035412 (2012).

\bibitem{LiuWM11} W. Wu, S. Rachel, W.-M. Liu, and K. Le Hur, arXiv:1106.0943.

\bibitem{LiTao12} T. Li, Euro. Phys. Lett. {\bf 97}, 37001 (2012).

\bibitem{Moller08} A. M$\ddot{o}$ller, U. L$\ddot{o}$w, T. Taetz, M. Kriener,
G. Andre, F. Damay, O. Heyer, M. Braden, and J. A. Mydosh, Phys.
Rev. B {\bf 78}, 024420 (2008).

\bibitem{Yehia10} M. Yehia, E. Vavilova, A. M$\ddot{o}$ller, T. Taetz,
U. L$\ddot{o}$w, R. Klingeler, V. Kataev, and B. B$\ddot{u}$chner,
Phys. Rev. {\bf B 81}, 060414(R) (2010).

\bibitem{Yan11} Y.-J. Yan, Z.-Y. Li, T. Zhang, X.-G. Luo, G.-J. Ye, Z.-J.
Xiang, P. Cheng, L.-J. Zou, and X. H. Chen, Phys.
Rev. {\bf B 85}, 085102 (2012).

\bibitem{WIEN2K} P. Blaha, K. Schwarz, G. Madsen, D. Kvasnicka, and J. Luitz:
"Computer Code WIEN2k, an augmented plane wave plus local orbitals
program for calculating crystal properties", Karlheinz
Schwarz, Technische Universit$\ddot{s}$t Wien, Austria, (2001).

\bibitem{PBE} J. P. Perdew, K. Burke, and M. Ernzerhof, Phys. Rev. Lett. {\bf 77},
3865 (1996).

\bibitem{PRB79-125201} M. Yashima, and R. O. Suzuki, Phys. Rev. B {\bf 79},
125201 (2009).

\bibitem{JPCM22-416002} H. Wang, and U. Schwingenschl$\ddot{o}$gl, J. Phys.
Condens. Matter {\bf 22}, 416002 (2010).

\bibitem{PRL104-047401} F. Rodolakis, P. Hansmann, J.-P. Rueff, A. Toschi,
M. W. Haverkort, G. Sangiovanni, A. Tanaka, T. Saha-Dasgupta, O. K. Andersen,
K. Held, M. Sikora, I. Alliot, J.-P. Iti$\acute{e}$, F. Baudelet, P. Wzietek,
P. Metcalf, and M. Marsi, Phys. Rev. Lett. {\bf 104}, 047401 (2010).

\bibitem{Kataev05} V. Kataev, A. M$\ddot{o}$ller, U. L$\ddot{o}$w, W. Jung, N.
Schittner, M. Kriener, and A. Freimuth, J. Magn. Magn. Mater. {\bf 290-291},
310 (2005).

\bibitem{Ran07} Y. Ran, M. Hermele, P. A. Lee, X.-G. Wen, Phys. Rev. Lett.
{\bf 98}, 117205 (2007).

\bibitem{low-D-AFM} K. Oka, I. Yamada, M. Azuma, S. Takeshlta, K. H. Satoh, A. Koda,
R. Kadono, M. Takano, and Y. Shlmakawa, Inorg. Chem, {\bf 47}, 7355 (2008).

\bibitem{Mezzacapo12} F. Mezzacapo and M. Boninsegni, Phys. Rev, {\bf 85}, 060402 (2012).





\noindent

\end{references}
\end{document}